\documentclass[final,twocolumn]{elsarticle}

\usepackage{textcomp}
\usepackage{color}
\usepackage{amssymb,amsmath}
\usepackage{mathrsfs}
\usepackage{float}
\usepackage{graphicx}

\biboptions{sort,compress}

\journal{Nuclear Instruments and Methods B}
\begin{document}
\begin{frontmatter}
\title{Ba-ion extraction from a high pressure Xe gas for double-beta decay studies with EXO}
\author[stanford]{T.~Brunner\corref{cor1}}
\cortext[cor1]{Corresponding Author}
\ead{tbrunner@stanford.edu}
\author[stanford]{D.~Fudenberg}
\author[stanford]{A.~Sabourov\fnref{fn1}}
\author[itep,gsi]{V.L.~Varentsov}
\author[stanford]{G.~Gratta}
\author[canada,triumf]{D.~Sinclair}
\author{for the EXO collaboration}
\address[stanford]{Dept. of Physics, Stanford University, Stanford, CA, USA}
\address[itep]{Institute for Theoretical and Experimental Physics, Moscow, Russia}
\address[gsi]{Facility for Antiproton and Ion Research in Europe (FAIR), Darmstadt, Germany }
\address[canada]{Dept. of Physics, Carleton University, Ottawa, ON, Canada}
\address[triumf]{TRIUMF, Vancouver, BC, Canada}
\fntext[fn1]{Current address: Orlando, FL, USA}

\begin{abstract}
An experimental setup is being developed to extract Ba ions from a high-pressure Xe gas environment. It aims to transport Ba ions from 10\,bar Xe to vacuum conditions. The setup utilizes a converging-diverging nozzle in combination with a radio-frequency (RF) funnel to move Ba ions into vacuum through the pressure drop of several orders of magnitude. This technique is intended to be used in a future multi-ton detector investigating double-beta decay in $^{136}$Xe. Efficient extraction and detection of Ba ions, the decay product of Xe, would allow for a background-free measurement of the $^{136}$Xe double-beta decay.
\end{abstract}

\end{frontmatter}
\section{Introduction}\label{intro} 
Several double-beta ($\beta\beta$) decay experiments are currently trying to determine the nature of neutrinos. An unambiguous observation of the lepton-number violating zero-neutrino $\beta\beta$ decay would require neutrinos to be Majorana particles. Furthermore, if the main contribution to this process is the exchange of light Majorana neutrinos one can extract the effective Majorana neutrino mass from the half life of the decay. However, half lives greater than $10^{25}$\,years pose great experimental challenges. In order to reach the sensitivities needed to observe such decays requires a significant increase in detector mass and observation time relative to existing experiments, while minimizing naturally occurring radioactive backgrounds.
Conventional detectors are limited by natural backgrounds thus the reachable sensitivity to the effective Majorana neutrino mass scales as $\left(N\ t\right)^{-1/4}$ with $N$ being the number of mother nuclei and $t$ being the observation time. In the case of a background-free experiment this situation is improved and the sensitivity scales with $\left(N\ t\right)^{-1/2}$ \cite{Dan00}. Of all the isotopes under consideration for $\beta\beta$-decay studies $^{136}$Xe is the only one that allows one to extract and identify the daughter of the decay immediately after the decay occurred \cite{Moe91}:
\begin{equation}
^{136}Xe \rightarrow\ ^{136}Ba^{++}+2e^-+\left(2\ \textnormal{or}\ 0\right)\ \bar{\nu_e}.
\end{equation}
A positive identification of the decay product $^{136}$Ba$^{++}$ allows the differentiation between $\beta\beta$ decays and natural background and hence a background-free measurement.

EXO-200, one of the experiments investigating the $\beta\beta$ decay of $^{136}$Xe, first determined the half life of $2\nu\beta\beta$ \cite{Ack11} and later published a limit for the half life of $0\nu\beta\beta$ in $^{136}$Xe \cite{Aug12}. In parallel to the operation of EXO-200 the collaboration is developing techniques to tag, i.e., extract from the volume and identify, Ba ions in both liquid and gaseous Xe (gXe). This work presents the concept and status of Ba-tagging endeavors in gXe, focusing on the extraction of Ba ions from a high-pressure Xe environment. 

\begin{figure}
  \centering
  \includegraphics[width=.45\textwidth]{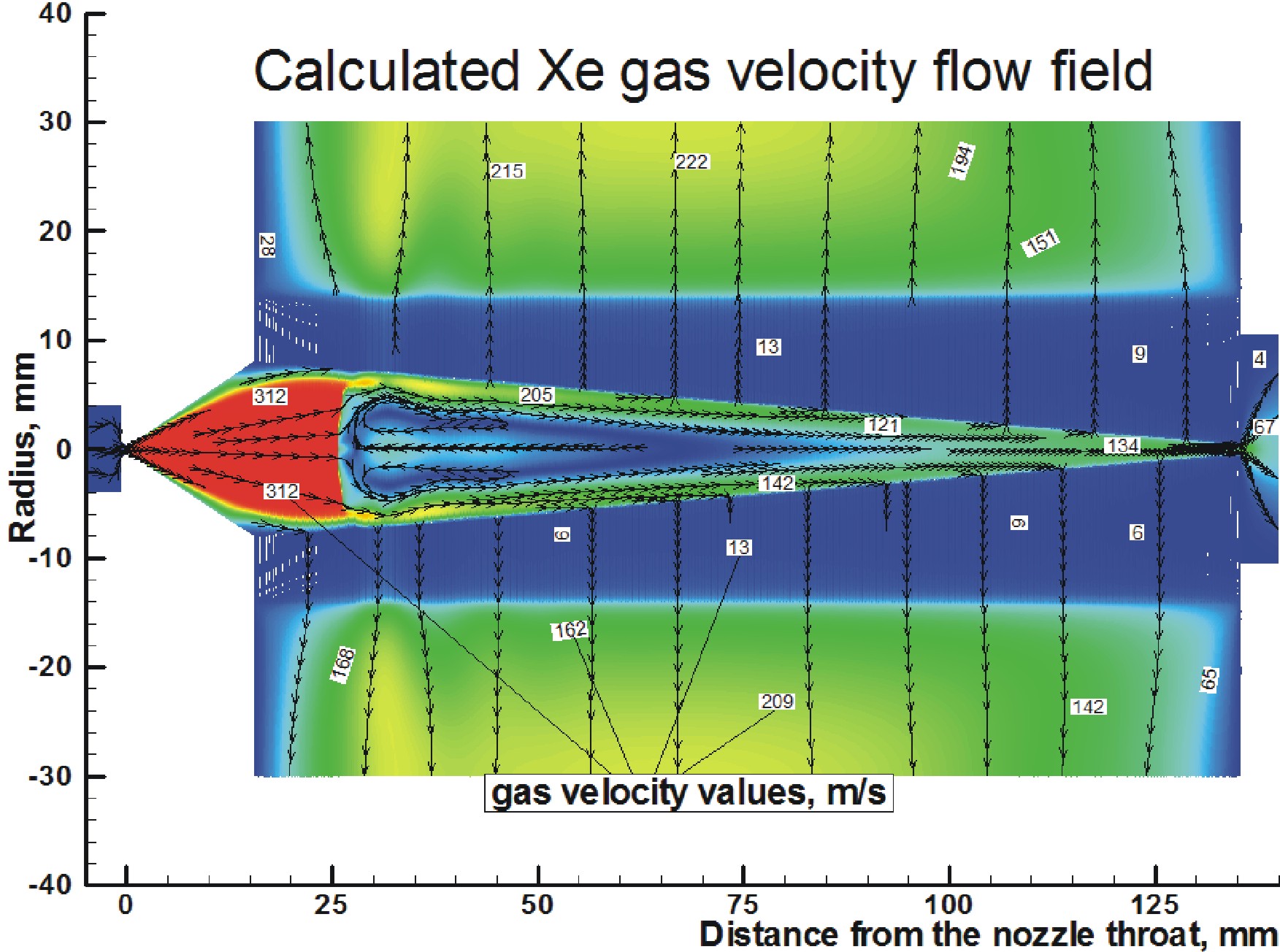}
  \caption{Calculated Xe-gas velocity flow field of the converging-diverging nozzle in combination with the electrode stack.}
  \label{fig:Simulation}
\end{figure}%
The general concept for a multi-ton gXe detector consists of a time-projection chamber (TPC) operated at 10\,bar enriched $^{136}$Xe. This records the energy and the position of events occurring in the active detector volume by conventional TPC techniques. If an event occurs within an energy window around $Q_{\beta\beta}(^{136}\textnormal{Xe})$, the electric field inside the TPC is modified to move ions from the decay volume to an exit port where they are flushed out of the TPC in a continuous flow of Xe gas. This Xe gas, carrying the extracted ions, is injected into a RF funnel through a converging-diverging nozzle. In the RF funnel,  Xe gas is removed through spaces in the funnel's electrode stack while Ba ions are confined by the applied RF field. Following the RF funnel Ba ions are transported to a downstream chamber filled with triethylamine (TEA). There, Ba$^{++}$ is converted to Ba$^{+}$ \cite{Sin11}. After the charge exchange process, the Ba ions are captured in a linear Paul trap \cite{Fla07} where they are identified by means of laser spectroscopy \cite{Gre07}.
\section{Ba$^+$ extraction from Xe gas - a prototype} 
At Stanford a setup has been developed to operate a 10\,bar natural Xe-gas jet with the ability to recycle Xe gas. The setup consists of the 10\,bar Xe chamber (referred to as chamber A) that is connected to a second vacuum chamber (referred to as chamber B) through a nozzle. The second vacuum vessel has a diameter of about 603\, mm and a cryo pump mounted in it. During gas-jet operation, Xe freezes to the cryo pump. At gas flow rates of 0.5\,g/s a base pressure in the mbar range is present in the chamber B. At this flow rate 1.5\,kg of Xe allows the operation of the jet at 10\,bar stagnation pressure in chamber A for about 30 min. Afterwards, the Xe-gas storage bottles are submerged in LN$_2$ while the cryo pump is warmed up. The Xe gas condenses in the storage gas bottles for reuse. A SAES purifier installed along the Xe supply line cleans contaminants from the Xe gas each time the gas jet is operated.  
\subsection{Concept of an RF funnel} 
An ion beam extraction system for gas cells based on the RF funnel has been suggested in 2001 \cite{Var01}. The operation of this system is described in details elsewhere (see e.g. Refs. \cite{Var02,Var04}). The RF-only ion funnel device, which consists of a stack of 301 electrodes, is placed on the converging-diverging nozzle axis in immediate vicinity of the nozzle exit plane. Ba ions are injected into the funnel via the supersonic Xe-gas jet. An RF voltage is applied with alternating phases delivered to neighboring electrodes. The funnel here works as a very good filter because the most part of the Xe-gas is evacuated through the gaps between funnel electrodes by pumping while the repelling force, which is generated by the RF field near the inner surface of the funnel electrodes and directed axially inwards, confines the Ba ions inside the funnel. At the same time the Xe-gas flow inside the funnel is strong enough to rapidly and very efficiently transport the Ba ions through the funnel into high vacuum conditions, eliminating the need for an additional electro-static drag potential. The geometry has been optimized in Monte-Carlo simulations for the operation in Stanford's Xe-gas recirculation setup described above. A simulated gas flow map is presented in Fig.\,\ref{fig:Simulation}. At an RF frequency of $f=2.6$\,MHz and an amplitude of V$_{\textnormal{PP}}$= 50\,V the simulated transport efficiency is 72.0(7)\%. Higher frequencies and amplitudes show simulated efficiencies of more than 95\%.
\begin{figure*}
	\centering
		\includegraphics[width=.75\textwidth]{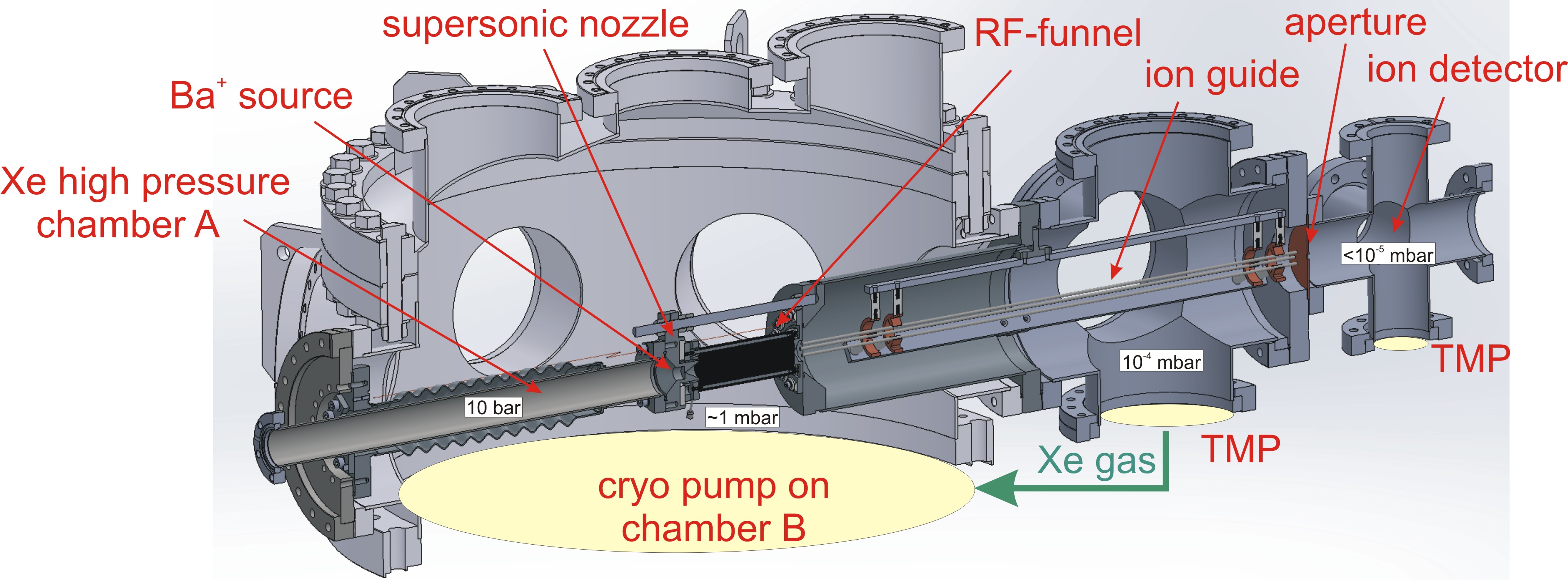}
	\caption{Section view of Stanford's gXe setup. Xe gas is injected from the left through the supersonic nozzle. The RF funnel is placed in the center of the chamber. Downstream of the funnel the ions are guided through a differential pumping region before they are detected by an ion detector.}
	\label{fig:whole-concept}
\end{figure*}
\subsection{The RF funnel} 
Based on the RF-funnel simulations a mechanical setup has been designed and built that mounts inside chamber B. A section view of the model is presented in Fig.\,\ref{fig:whole-concept}. The RF funnel is placed at the center of chamber B above the cryo pump. Xe gas is injected through the nozzle that is mounted on the high pressure chamber A. Downstream of the funnel an ion guide transports the ions through a differential pumping section before they are detected with an ion detector. This is an intermediate step to test the performance of the funnel.

The converging-diverging nozzle was machined in a special CF2.75$''$ flange through electrode-spark discharge machining\footnote{EDN Labs Ltd., www.edmlabsltd.com}. The half-angles of subsonic and supersonic part are $45^{\circ}$ and $26.6^{\circ}$, respectively. The lengths of subsonic and supersonic part are 0.5\,mm and 15.5\,mm, respectively. The throat diameter is 0.3\,mm and the exit diameter is 16.0\,mm. 

\begin{figure}
  \centering
  \includegraphics[trim = 230 120 905 40, clip, angle=90, width=.45\textwidth]{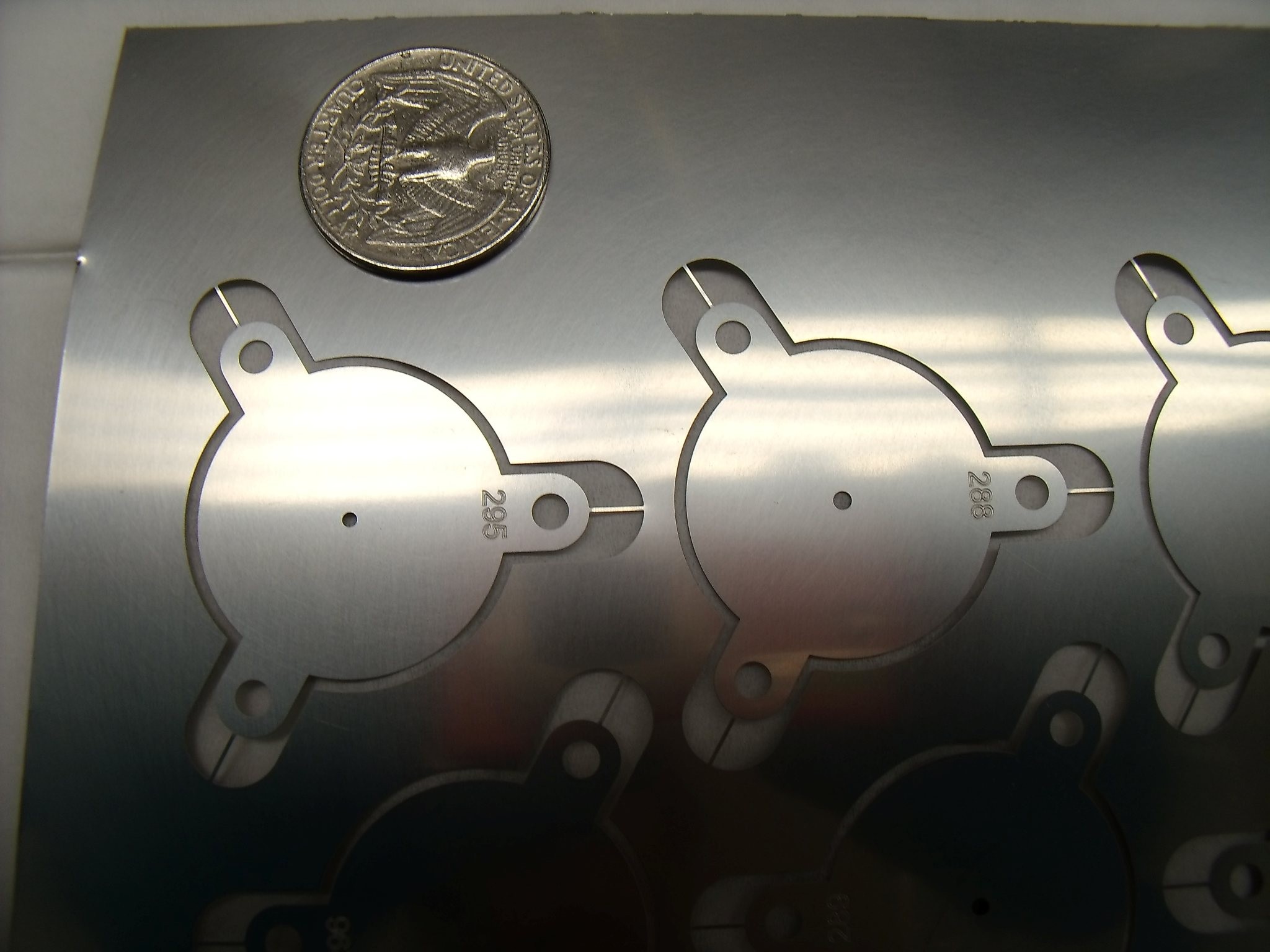}
  \caption{Photo of electrode \#295 (Quarter for scale). The electrodes were photo-etched in a stainless-steel sheet and numbered individually. Prior to its assembly the electrode was held in place by `legs' extending from the three equally spaced latches used to mount the electrode. These legs are cut off during installation.}
  \label{fig:electrode}
\end{figure}

The 301 funnel electrodes were photo-etched\footnote{Newcut Inc. New York} in a 0.1016 ($\pm$0.0025)\,mm thick stainless steel sheet\footnote{305\,mm x 305\,mm alloy 316, ESPI Metals} with manufacturing tolerances better than 0.025\,mm \cite{Eng12}. The outer diameter of each ring electrode is 28\,mm. Each electrode has an aperture etched at the center. The aperture diameter decreases from 16.0\,mm to 1\,mm in steps of 0.05\,mm per electrode. Three mounting latches with holes are equally spaced on the electrode. Each electrode has its number etched into it to simplify the assembly process. A picture of electrode \# 295 is shown in Fig.\,\ref{fig:electrode}. The electrodes were held in place in the metal sheet by three small 'legs'. For the installation these legs were cut close to the electrode.

\begin{figure}
  \centering
  \includegraphics[width=.45\textwidth]{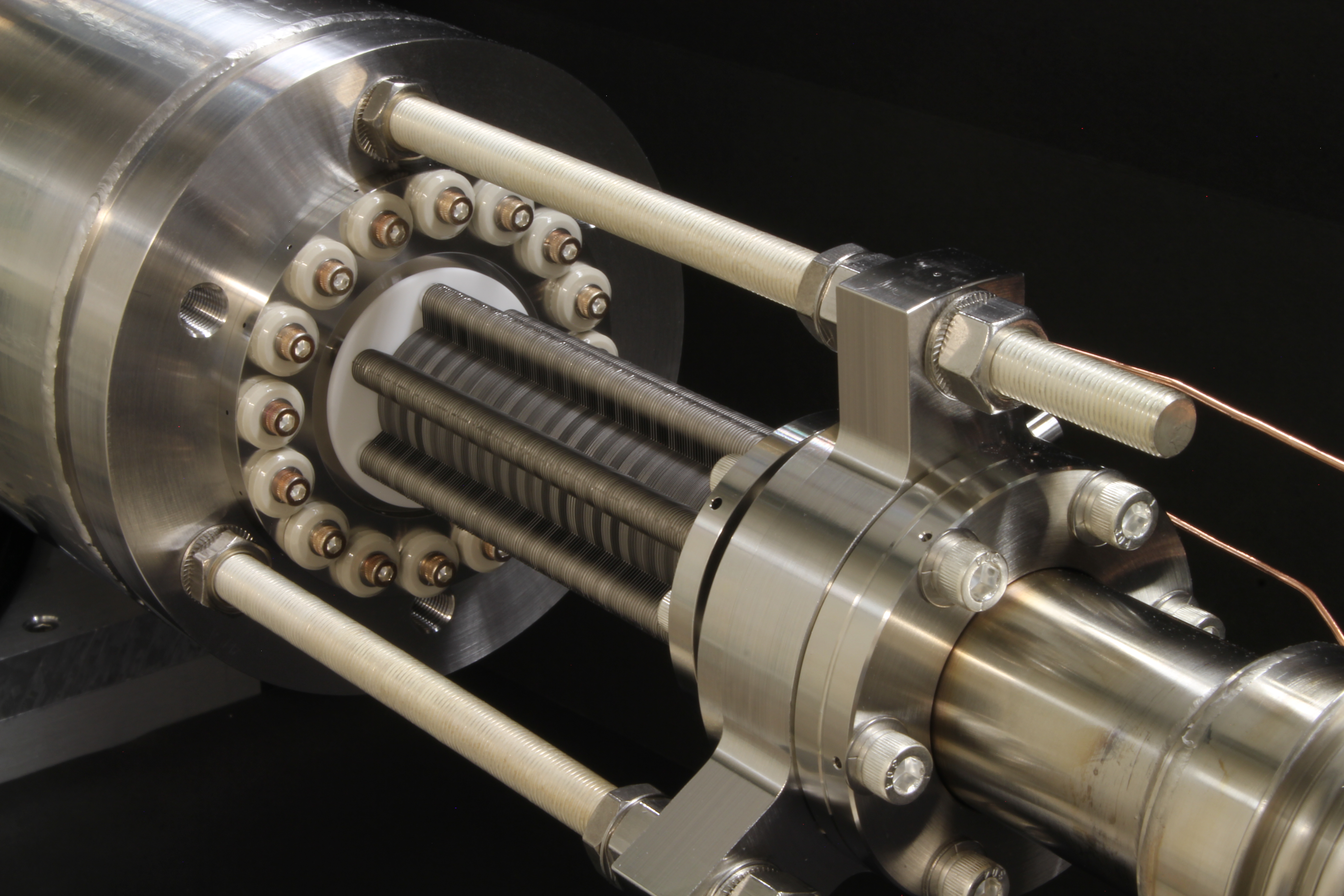}
  \caption{Photo of the RF funnel installed in its final configuration. The high-pressure Xe will be injected from the right side, pass through the nozzle, then the RF funnel, and proceed into the vacuum chamber on the left, where  the downstream ion guide will be installed.}
  \label{fig:FunnelPic}
\end{figure}

The electrodes were stacked alternating between two sets of three mounting rods. Both sets are electrically insulated. Within one stack the electrodes are spaced by 0.6096(+0.0051/-0.0102)\,mm thick stainless steel spacers\footnote{203\,mm x 305\,mm alloy 316, ESPI Metals} that were photo-etched. This results in a distance of 0.25\,mm between the faces of two neighboring electrodes in the stack.
The RF funnel stack was fixed to the nozzle, which was then mounted on the downstream vacuum chamber by three 3/8$''$ threaded rods. The funnel was mounted in such a way that the exit aperture of the funnel is concentric with the entrance hole of the downstream vacuum chamber. The distance between electrode \#301 and downstream vacuum chamber is 0.2388\,mm. A picture of high-pressure chamber, nozzle and funnel mounted on the downstream chamber is presented in Fig.\,\ref{fig:FunnelPic}. 
The capacitance of this assembly outside of chamber B is 6.055\,nF, resulting in a resonance frequency close to 2.6\,MHz. At this frequency amplitudes up to 80\,V$_{PP}$ were successfully applied to the funnel. In a separate test, N$_2$ gas of up to 12\,bar was applied in chamber A, resulting in a gas jet on the funnel. During testing no electrical short between the stacks was observed due to electrode movement.
\section{Outlook} 
An RF-funnel setup to extract Ba$^{+(+)}$ from a 10\,bar Xe gas jet is currently being designed and constructed. The RF-funnel is fully assembled and installed inside chamber B. The initial RF tests will be repeated with and without Xe gas jet operation. A Gd driven Ba-ion source \cite{Mon10} is currently being fabricated. This source will be placed right at the nozzle as shown in Fig.\,\ref{fig:whole-concept}. For the downstream ion guide a sextupole-ion guide is under development. The geometry of this guide is being optimized in SimIon simulations \cite{Dah00}. First ion extractions are planned for summer 2013.
\section{Acknowledgments}
The Ba-tagging in Xe gas project is funded by NSF. We thank K. Merkle, J. Kirk, and R. Conley for their help in machining parts. 
We also thank M. Brodeur, T. Dickel, J. Dilling, M. Good, R. Ringle, P. Schury, S. Schwarz and K. Skarpaas for fruitful discussions and their support of this project.

\bibliographystyle{elsarticle-num}
\bibliography{bibliography}
\end{document}